\documentclass[aps,11pt,nofootinbib,preprintnumbers,groupedaddress,superscriptaddress]{revtex4-2}

\usepackage{amssymb, amsfonts, amsmath, bm}
\usepackage[
]{graphicx}
\usepackage{color}
\usepackage{mathrsfs}
\usepackage{cases}

\usepackage[
]{hyperref}
\hypersetup{%
 colorlinks=true,
  linkcolor=blue,   
  citecolor=blue,   
  urlcolor=blue
}
\newcommand{\orcid}[1]{%
  \href{https://orcid.org/#1}{\includegraphics[height=0.7em]{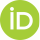}}%
}

\begin{document}
\title{{\large Deflection angle in the strong deflection limit:\\ A perspective from local geometrical invariants and matter distributions}}
\author{Takahisa Igata\:\!\orcid{0000-0002-3344-9045}
}
\email{takahisa.igata@gakushuin.ac.jp}
\affiliation{
Department of Physics, Gakushuin University,\\
Mejiro, Toshima, Tokyo 171-8588, Japan}
\date{\today}

\begin{abstract}
In static, spherically symmetric spacetimes, the deflection angle of photons in the strong deflection limit exhibits a logarithmic divergence. We introduce an analytical framework that clarifies the physical origin of this divergence by employing local, coordinate-invariant geometric quantities alongside the properties of the matter distribution. In contrast to conventional formulations---where the divergence rate $\bar{a}$ is expressed via coordinate-dependent metric functions---our approach relates $\bar{a}$ to the components of the Einstein tensor in an orthonormal basis adapted to the spacetime symmetry. By applying the Einstein equations, we derive the expression
\begin{align*}
\bar{a}=\frac{1}{\sqrt{1-8\pi R_{\mathrm{m}}^2\left(\rho_{\mathrm{m}}+\Pi_{\mathrm{m}}\right)}},
\end{align*}
where $\rho_{\mathrm{m}}$ and $\Pi_{\mathrm{m}}$ denote the local energy density and tangential pressure evaluated at the photon sphere of areal radius $R_{\mathrm{m}}$. This result reveals that $\bar{a}$ is intrinsically governed by the local matter distribution, with the universal value $\bar{a}=1$ emerging when $\rho_{\mathrm{m}}+\Pi_{\mathrm{m}}=0$. Notably, this finding resolves the long-standing puzzle of obtaining $\bar{a}=1$ in a class of spacetimes supported by a massless scalar field. Furthermore, these local properties are reflected in the frequencies of quasinormal modes, suggesting a profound connection between strong gravitational lensing and the dynamical response of gravitational wave signals. 
\end{abstract}

\maketitle

\newpage

\section{Introduction}
\label{sec:1}
Recent advances in observational astronomy, including the direct imaging of shadows cast by supermassive compact objects at the centers of galaxies~\cite{EventHorizonTelescope:2019dse,EventHorizonTelescope:2022wkp}, provide strong evidence for the existence of photon spheres in strong gravitational fields. A photon sphere, theoretically predicted to occur near various compact objects such as black holes, is defined by the presence of unstable photon circular orbits~\cite{Claudel:2000yi}. This feature plays a crucial role in gravitational lensing within the strong deflection regime, resulting in observable phenomena such as multiple images, higher-order relativistic rings, and shadows. The light deflection angle, which exhibits a logarithmic divergence near the photon sphere, encapsulates these phenomena and provides crucial insights into the physical properties of the region, as first demonstrated in the Schwarzschild spacetime~\cite{Darwin:1959}. Therefore, a thorough investigation of this divergent behavior enhances our understanding of the local spacetime geometry around photon spheres and establishes a foundation for rigorous tests of gravitational theories.

Building on these insights, numerous studies have focused on developing analytical formulations for the deflection angle in the strong deflection limit and exploring their applications. An explicit expression for the deflection angle in this context was first derived in Ref.~\cite{Bozza:2002zj}, establishing the foundation for subsequent research. Later, this approach was refined in Refs.~\cite{Tsukamoto:2016jzh,Shaikh:2019itn}, thus extending its applicability to various spacetimes. Because of significant astrophysical interest in this regime and the methodological simplicity of these techniques, they have been applied to a broad range of compact astrophysical objects. Notable examples include the Schwarzschild black hole~\cite{Bozza:2001xd}, the Reissner--Nordstr\"om black hole~\cite{Eiroa:2002mk, Tsukamoto:2016oca}, regular black holes~\cite{Eiroa:2010wm}, naked singularities~\cite{Chen:2023uuy,Sarkar:2006ry}, wormholes~\cite{Tsukamoto:2016qro, Nandi:2016uzg,Shaikh:2019jfr, Nandi:2006ds, TejeiroS:2005ltc, Bhattacharya:2019kkb,Izmailov:2019tyq}, ultracompact objects~\cite{Shaikh:2019itn}, gravastars~\cite{Kubo:2016ada}, and other black holes~\cite{Soares:2024rhp,Chakraborty:2016lxo}. Additionally, formulations in the strong deflection limit have been extended to incorporate finite-distance corrections~\cite{Ishihara:2016sfv,Takizawa:2021gdp} and to describe the deflection of massive particles~\cite{Feleppa:2024kio}. Consequently, the logarithmic divergence of the deflection angle in this regime has been consistently verified across a wide range of spacetimes. 

Furthermore, a fundamental relationship exists between the dynamics of photons around photon spheres and the quasinormal mode (QNM) frequencies of compact objects in the eikonal limit~\cite{Cardoso:2008bp}. Specifically, the rate of logarithmic divergence of the deflection angle is related to the imaginary part of the QNM frequencies~\cite{Stefanov:2010xz,Raffaelli:2014ola}. This connection implies that the strong deflection limit also reflects the inherent dynamical instability of photon orbits, as characterized by the corresponding Lyapunov exponent. Consequently, observations in the strong deflection regime can complement gravitational wave measurements by providing a multifaceted probe of the fundamental properties of compact objects and by serving as a powerful tool for testing gravitational theories.

Previous studies on the deflection angle in the strong deflection limit have identified the integral accounting for the logarithmic divergence, thereby evaluating the strong field limit coefficients $\bar{a}$ and $\bar{b}$, which determine the divergence rate and the constant offset correction, respectively. As explicitly demonstrated in this paper, the deflection angle is given by
\begin{align}
\alpha(b) =-\bar{a} \log \left(
\frac{b}{b_{\mathrm{c}}}-1
\right)+\bar{b}+O\left(\left(
\frac{b}{b_{\mathrm{c}}}-1
\right)^{1/2}
\log\left(
\frac{b}{b_{\mathrm{c}}}-1
\right)
\right).
\end{align}
Here, $b$ denotes the impact parameter, and $b_{\mathrm{c}}$ is its critical value. This formula has been successfully applied to a wide variety of astrophysical scenarios. However, two fundamental issues remain unresolved in the current formulation. First, the physical interpretation of $\bar{a}$ and $\bar{b}$ is not yet fully understood. Although these coefficients are expressed in terms of the local values of metric functions and their derivatives (and are therefore practically useful), their precise relationship to physical quantities---such as the local spacetime geometry or the influence of matter fields---remains unclear. Second, these coefficients, in their current form, are coordinate dependent. Since they are related to observable quantities, it is essential to reformulate them in a coordinate-independent manner in accordance with the principles of relativity, thus enabling a more universal physical interpretation.

To address these issues, we establish a direct relationship between the deflection angle in the strong deflection regime and local, coordinate-invariant physical quantities. Our formulation links the strong field limit coefficients to the properties of the photon sphere, thereby clarifying their role in photon dynamics. By expressing these coefficients in terms of the tetrad components of the Einstein tensor, we derive local, coordinate-independent expressions. Furthermore, by applying the Einstein equation, we relate the curvature to matter fields (energy density and pressure) and reveal a fundamental connection between the strong field coefficients and the underlying matter content across various spacetimes. This approach facilitates direct observational comparisons and deepens our understanding of the fundamental origins of the deflection angle.

This paper is organized as follows. In Sec.~\ref{sec:2}, we review photon motion and the conditions for circular orbits in static, spherically symmetric spacetimes. We introduce the general metric, define the Misner--Sharp mass, and derive the photon equations of motion to establish the criteria for unstable photon circular orbits. In Sec.~\ref{sec:3}, we derive the deflection angle in the strong deflection limit by expanding the integral near the photon sphere and isolating the logarithmic divergence. In Sec.~\ref{sec:4}, we express our results in terms of the tetrad components of the Einstein tensor to provide a coordinate-invariant description of the local spacetime geometry. In Sec.~\ref{sec:5}, we relate the strong field limit coefficients to local matter fields, namely, energy density, radial pressure, and tangential pressure, and discuss their implications for several spacetime models. Finally, in Sec.~\ref{sec:6}, we summarize our findings, discuss their implications for gravitational lensing and tests of gravitational theories, and suggest directions for future work.

Throughout this paper, we use geometrized units in which the gravitational constant $G$ and the speed of light $c$ are set to unity (i.e., $G=1$ and $c=1$).

\section{Photon dynamics and circular orbits in static, spherically symmetric spacetimes}
\label{sec:2}
We begin with the most general line element
of a static, spherically symmetric spacetime, given by
\begin{align}
\mathrm{d}s^2=-e^{\nu(r)}\:\!\mathrm{d}t^2+e^{\lambda(r)}\:\!\mathrm{d}r^2+R(r)^2(\mathrm{d}\theta^2+\sin^2\theta\:\!\mathrm{d}\varphi^2),
\label{eq:metric}
\end{align}
where $t$, $r$, and $(\theta,\varphi)$ denote the time, radial, and angular coordinates, respectively. The functions $\nu(r)$, $\lambda(r)$, and $R(r)$ are arbitrary; however, one of these can be fixed by an appropriate gauge choice. Here, $R(r)$ represents the areal radius [i.e., $R(r)>0$]. We assume that the spacetime is asymptotically flat, meaning that as $R\to \infty$ (corresponding to $r\to r_{\infty}$, with $r_{\infty}$ denoting the coordinate value at spatial infinity), the metric functions approach $e^{\nu}\to 1$ and $e^{\lambda}\to (R')^2$, ensuring convergence to the Minkowski metric,%
\footnote{In our formulation, we do not impose the coordinate condition that the radial coordinate asymptotically approaches the areal radius at infinity, as is commonly done in the literature (e.g., Refs.~\cite{Bozza:2002zj,Tsukamoto:2016jzh}).} 
where, and hereafter, the prime denotes differentiation with respect to $r$.

We introduce the Misner--Sharp mass, which represents the gravitational mass enclosed within a sphere of radius $r$~\cite{Misner:1964je,Hayward:1994bu,Kinoshita:2024wyr}. It is defined by 
\begin{align}
m(r)=\frac{R}{2}\left[\:\!1-g^{ab} (\nabla_a R)(\nabla_b R) \:\!\right],
\label{eq:mdef}
\end{align}
where $g^{ab}$ denotes the inverse metric and $\nabla_a$ denotes the covariant derivative. For the metric \eqref{eq:metric}, Eq.~\eqref{eq:mdef} reduces to
\begin{align}
m(r)=\frac{R}{2}\left[1 - e^{-\lambda} (R')^2\right],
\end{align}
where a prime denotes differentiation with respect to the radial coordinate $r$.
Furthermore, if $R'\neq 0$, the metric function $e^{\lambda}$ can be expressed in terms of $m(r)$ as
\begin{align}
e^{\lambda(r)} = \frac{(R')^2}{1-2m(r)/R}\, , 
\label{eq:m}
\end{align}
provided that $R>2m$, which ensures that the metric is well defined.

Next, we examine photon dynamics in this spacetime. 
We assume that photons propagate along null geodesics, that is, they interact only gravitationally with the background spacetime. This standard assumption is consistent with the conventional treatments of strong gravitational lensing in curved spacetimes (e.g., Refs.~\cite{Bozza:2002zj,Tsukamoto:2016jzh}).
In situations where light-matter interactions are non-negligible, such as in refractive or dispersive media (see, e.g., Refs.~\cite{Tsupko:2013cqa, Perlick:2015vta}), the photon dynamics deviate from null geodesics, and the present formulation would need to be generalized accordingly.

Because of the spherical symmetry of the metric, their orbits can be confined to the equatorial plane (i.e., $\theta=\pi/2$) without loss of generality. Under these assumptions, the Lagrangian for a photon is given by
\begin{align}
\mathscr{L} = \frac{1}{2}\left(-e^{\nu}\:\!\dot{t}^2 + e^{\lambda}\:\!\dot{r}^2 + R^2\:\!\dot{\varphi}^2\right),
\end{align}
where the overdot denotes differentiation with respect to an affine parameter along the null geodesic. Since $\mathscr{L}$ is independent of $t$ and $\varphi$, their corresponding conjugate momenta are conserved. We identify these conserved quantities as the photon's energy and angular momentum:
\begin{align}
E &= e^{\nu}\:\!\dot{t}, \\
L &= R^2\dot{\varphi}.
\end{align}
The null condition ($\mathscr{L}=0$) then yields the first-order radial differential equation:
\begin{align}
\dot{r}^2 + e^{-\lambda}\left(\frac{L^2}{R^2}-E^2e^{-\nu}\right)=0.
\label{eq:radialeq}
\end{align}
Dividing this equation by $\dot{\varphi}^2$ (assuming $\dot{\varphi}\neq0$) leads to the orbital differential equation
\begin{align}
\left(\frac{\mathrm{d}r}{\mathrm{d}\varphi}\right)^2 + V(r) &= 0,
\label{eq:orbitaleq}
\end{align}
with the effective potential defined as
\begin{align}
V(r) = R^2e^{-\lambda}\left(1-\frac{R^2e^{-\nu}}{b^2}\right),
\end{align}
where the impact parameter is given by $b=L/E$.

For a photon circular orbit, both $\dot{r}$ and $\ddot{r}$ must vanish. This requires $V = 0$ and $V' = 0$. Setting $V=0$ immediately gives
\begin{align}
b^2 = R^2 e^{-\nu}.
\label{eq:bsq}
\end{align}
Combining this with $V'=0$ yields
\begin{align}
\nu' = \frac{2R'}{R}.
\label{eq:PSradius}
\end{align}
Let $r_{\mathrm{m}}$ denote the radius where these conditions hold. Then, the circular orbit occurs at $r=r_{\mathrm{m}}$, and the corresponding critical impact parameter is 
\begin{align}
b_{\mathrm{c}}^2=R_{\mathrm{m}}^2 e^{-\nu_{\mathrm{m}}}, 
\label{eq:bc}
\end{align}
where $R_{\mathrm{m}}\equiv R(r_{\mathrm{m}})$ and $\nu_{\mathrm{m}}\equiv \nu(r_{\mathrm{m}})$. 
The stability of the circular orbit is determined by the second derivative of the effective potential. Evaluating $V''$ at $r=r_{\mathrm{m}}$ for $b=b_{\mathrm{c}}$ yields
\begin{align}
V''_{\mathrm{m}}=e^{-\lambda_{\mathrm{m}}}\left[\:\!
\nu''_{\mathrm{m}}R_{\mathrm{m}}^2 +2(R'_{\mathrm{m}})^2-2R_{\mathrm{m}} R''_{\mathrm{m}}
\:\!\right],
\label{eq:V''m}
\end{align}
where $\lambda_{\mathrm{m}}\equiv \lambda(r_{\mathrm{m}})$, 
$R'_{\mathrm{m}}\equiv R'(r_{\mathrm{m}})$, and $R''_{\mathrm{m}}\equiv R''(r_{\mathrm{m}})$. 
In our analysis, we focus on unstable photon circular orbits. Therefore, we require that the effective potential exhibits a local maximum at $r=r_{\mathrm{m}}$, i.e., $V''_{\mathrm{m}}<0$, or equivalently, 
\begin{align}
\nu''_{\mathrm{m}}R_{\mathrm{m}}^2 +2(R'_{\mathrm{m}})^2-2R_{\mathrm{m}} R''_{\mathrm{m}}<0. 
\end{align}
Consequently, the constant-$r$ timelike hypersurface at $r=r_{\mathrm{m}}$ defines the photon sphere.

\section{Derivation of the deflection angle in the strong deflection limit}
\label{sec:3}

To quantify the bending of light in this spacetime, we define the deflection angle. Using the orbital differential equation~\eqref{eq:orbitaleq}, the total change in the photon's azimuthal angle, as it travels from infinity to the point of closest approach $r_0$ and then back to infinity, is given by 
\begin{align}
I(r_0)=2 \int_{r_0}^{r_\infty} \frac{\mathrm{d}r}{\sqrt{-V}}.
\end{align}
Since $V(r_0)=0$, the impact parameter $b$ is given by
\begin{align}
b^2=R_{0}^2\:\! e^{-\nu_0},
\label{eq:b0}
\end{align}
where $R_0\equiv R(r_0)$ and $\nu_0\equiv \nu(r_0)$.
Thus, the deflection angle is defined as
\begin{align}
\alpha(r_0) = I(r_0) - \pi.
\end{align}

Following Ref.~\cite{Bozza:2002zj}, we evaluate the deflection angle in the strong deflection limit. We introduce the new variable, 
\begin{align}
z=1-\frac{R_0}{R},
\end{align}
which differs from the variable used in the literature. This definition ensures that the final expressions are formulated primarily in terms of the areal radius. As a result, the formulation is both coordinate invariant and geometrically meaningful. With this substitution, the integral $I(r_0)$ becomes 
\begin{align}
I(r_0)=2 \int_0^1 
\frac{\mathrm{d}z}{\sqrt{-\frac{(R')^2}{R_0^2}(1-z)^4\:\!V}},
\label{eq:Iz}
\end{align}
assuming that $R'\neq 0$. 
Expanding $V$ and $(R')^2$ around $z=0$ (i.e., $r=r_0$ or $R=R_0$) up to subleading order, we obtain
\begin{align}
V &= \frac{R_0 V'_0}{R_0'} \:\! z + \left[\:\!
\left(\frac{R_0}{R_0'}-\frac{R_0^2R_0''}{2(R_0')^3}\right)V_0'
+\frac{R_0^2 }{2(R_0')^2}V_0''
\:\!\right]z^2+ O(z^3),
\\
(R')^2&=(R'_0)^2+2R_0R_0'' z+O(z^2),
\end{align}
where we denote $V'_0\equiv V'(r_0)$, $V''_0\equiv V''(r_0)$, and $R''_0\equiv R''(r_0)$. Defining
\begin{align}
c_1=-\frac{R'_0}{R_0} \:\!V'_0,
\quad
c_2=3\left(\frac{R_0'}{R_0}-\frac{R_0''}{2R_0'}\right)V_0'-\frac{V''_0}{2},
\end{align}
we find that, up to second order in $z$, the expression under the square root in Eq.~\eqref{eq:Iz} behaves as $c_1 z+c_2 z^2$. Accordingly, we define the integral 
\begin{align}
I_{\mathrm{D}}(r_0)=2\int_0^1 \frac{\mathrm{d}z}{\sqrt{c_1 z+c_2 z^2}}. 
\label{eq:divpart}
\end{align}
In the strong deflection limit (i.e., as $r_0\to r_{\mathrm{m}}$), we have $V'(r_0)\to 0$, which implies that $c_1\to 0$. Consequently, the integrand behaves as $(\sqrt{c_2}z)^{-1}$, leading to a logarithmic divergence as $z\to 0$. Hence, $I_{\mathrm{D}}(r_0)$ represents the divergent contribution to $I(r_0)$. The remaining finite part is then defined as
\begin{align}
I_{\mathrm{R}}(r_0)=I(r_0)-I_{\mathrm{D}}(r_0). 
\end{align}
Furthermore, integration of Eq.~\eqref{eq:divpart} yields an explicit expression for the divergent part:
\begin{align}
I_{\mathrm{D}}(r_0)=\frac{4}{\sqrt{c_2}} \log \frac{\sqrt{c_1+c_2}+\sqrt{c_2}}{\sqrt{c_1}}. 
\end{align}

To relate the deflection angle to the impact parameter $b$, we expand Eq.~\eqref{eq:b0} around the photon sphere, i.e., $r_0=r_{\mathrm{m}}$. Then, we obtain
\begin{align}
b=b_{\mathrm{c}} \left[\:\!
1-\frac{e^{\lambda_{\mathrm{m}}} V''_{\mathrm{m}}}{4 R_{\mathrm{m}}^2} (r_0-r_{\mathrm{m}})^2+O\left(\Big(\frac{
r_0}{r_{\mathrm{m}}}-1\Big)^3\right)
\:\!\right],
\label{eq:bexp}
\end{align}
where $b_{\mathrm{c}}$ is defined in Eq.~\eqref{eq:bc}. 

\subsection{$R'_{\mathrm{m}}\neq 0$}
We assume that $R'_{\mathrm{m}}\neq 0$. Under this assumption, we expand the coefficients $c_1$ and $c_2$ about $r_0=r_{\mathrm{m}}$ as
\begin{align}
c_1&=
-\frac{R_{\mathrm{m}}'V''_{\mathrm{m}}}{R_{\mathrm{m}}}
(r_0-r_{\mathrm{m}})+O\left(\Big(\frac{
r_0}{r_{\mathrm{m}}}-1\Big)^2\right),
\\
c_2&=-\frac{V''_{\mathrm{m}}}{2}+O\left(\frac{
r_0}{r_{\mathrm{m}}}-1\right).
\end{align}
By inverting Eq.~\eqref{eq:bexp}, we can express the coefficients $c_1$ and $c_2$ in terms of $(b/b_{\mathrm{c}}-1)$. This inversion yields
\begin{align}
c_1&=2 e^{-\lambda_{\mathrm{m}}/2}R'_{\mathrm{m}} \sqrt{-V_{\mathrm{m}}''}\left(
\frac{b}{b_{\mathrm{c}}}-1
\right)^{1/2}
+O\left(
\frac{b}{b_{\mathrm{c}}}-1
\right),
\\
c_2&=-\frac{V_{\mathrm{m}}''}{2}+O\left(
\Big(\frac{b}{b_{\mathrm{c}}}-1\Big)^{1/2}
\right).
\end{align}
These expressions lead directly to the following form for the divergent part of the deflection integral in terms of $b$:
\begin{align}
I_{\mathrm{D}}(r_0)=-\sqrt{-\frac{2}{V''_{\mathrm{m}}}} \log\left(
\frac{b}{b_{\mathrm{c}}}-1
\right)+\sqrt{-\frac{2}{V''_{\mathrm{m}}}} \log\left[\:\!
-\frac{e^{\lambda_{\mathrm{m}}}V''_{\mathrm{m}}}{(R'_{\mathrm{m}})^2}\:\!\right]
+O\left(\left(
\frac{b}{b_{\mathrm{c}}}-1
\right)^{1/2}
\log\left(
\frac{b}{b_{\mathrm{c}}}-1
\right)
\right).
\end{align}
Thus, the deflection angle in the strong deflection limit can be written as 
\begin{align}
\alpha(b) =-\bar{a} \log \left(
\frac{b}{b_{\mathrm{c}}}-1
\right)+\bar{b}+O\left(\left(
\frac{b}{b_{\mathrm{c}}}-1
\right)^{1/2}
\log\left(
\frac{b}{b_{\mathrm{c}}}-1
\right)
\right),
\label{eq:alphab}
\end{align}
with the strong field limit coefficients defined by
\begin{align}
\bar{a}&=\sqrt{-\frac{2}{V''_{\mathrm{m}}}},
\label{eq:aV''}
\\
\bar{b}&=-\bar{a} \log\left[\:\!
\frac{\bar{a}^2}{2}\left(1-\frac{2m(r_{\mathrm{m}})}{R_{\mathrm{m}}}\right)
\:\!\right]+ I_{\mathrm{R}}(r_\mathrm{m}) - \pi.
\end{align}
This expression demonstrates that the deflection angle diverges logarithmically as $b\to b_{\mathrm{c}}$. In particular, the coefficient $\bar{a}$ governs the rate of this logarithmic divergence and depends solely on $V''_{\mathrm{m}}$, i.e., the local curvature of the effective potential at the photon sphere. Since this divergence is an observable, local effect near the photon sphere, $\bar{a}$ is necessarily a coordinate-invariant and local quantity. In the following section, we show that $\bar{a}$, or equivalently, $V''_{\mathrm{m}}$, is connected to coordinate-invariant geometrical quantities. Additionally, it should be noted that this method completely isolates the divergent part of the integral and we can verify that the expression for $\bar{a}$ in terms of metric functions in the literature also reduces to Eq.~\eqref{eq:aV''}. This confirms that the extraction of the divergent behavior is independent of the choice of the expansion parameter $z$. 

In contrast, the coefficient $\bar{b}$ represents the constant offset correction to the deflection angle.%
\footnote{The notation $\bar{b}$ may be mistaken for the impact parameter; however, we adopt it here in accordance with established conventions in the literature.}
It incorporates contributions from both local and global properties of the spacetime. In particular, the local contributions arise through $V''_{\mathrm{m}}$ and $R_{\mathrm{m}}$ at the photon sphere, whereas the global contributions are given by $I_{\mathrm{R}}(r_{\mathrm{m}})$ and the quasi-local mass $m(r_{\mathrm{m}})$. Note that the decomposition of $\bar{b}$ into the term $I_{\mathrm{R}}(r_\mathrm{m}) - \pi$ and the other term depends on the choice of $z$. 
 
While previous works in the literature have expressed these coefficients arising from $I_{\mathrm{D}}$ in terms of metric functions, our formulation instead recasts them in terms of local quantities at the photon sphere (i.e., $V''_{\mathrm{m}}$ and $R_{\mathrm{m}}$) together with a well-defined quasi-local quantity [i.e., $m(r_{\mathrm{m}})$]. This approach not only clarifies the physical origin of the logarithmic divergence in the deflection angle but also identifies the observable quantities governing the phenomenon in the strong deflection limit.

It should be noted that if the coefficient of the $O(b/b_{\mathrm{c}}-1)^{1/2}$ term in $c_2$ is nonzero, then the error term in Eq.~\eqref{eq:alphab} generally appears at order $O((
b/b_{\mathrm{c}}-1
)^{1/2}
\log(
b/b_{\mathrm{c}}-1
)
)$ in any formulation that isolates the divergent integral. Although the evaluation of the regular integral $I_{\mathrm{R}}(r_{\mathrm{m}})$ is beyond the scope of the present work, detailed discussions on its evaluation can be found in Ref.~\cite{Bozza:2002zj}.

\subsection{$R'_{\mathrm{m}}=0$}
Next, we assume that $R'_{\mathrm{m}}=0$, which implies that the spacetime exhibits a wormhole geometry with the photon sphere located at the throat. In this case, the expansion of $c_1$ and $c_2$ around $r_0=r_{\mathrm{m}}$ reduces to%
\footnote{Although the first-order term in $(r_0/r_{\mathrm{m}}-1)$ in the expansion of $c_2$ does not affect the strong field limit coefficients, we retain it to capture the unique behavior of the higher-order correction terms for the Ellis--Bronnikov wormhole.}
\begin{align}
c_1&=
-\frac{R_{\mathrm{m}}''V''_{\mathrm{m}}}{R_{\mathrm{m}}}
(r_0-r_{\mathrm{m}})^2+O\left(\Big(\frac{r_0}{r_{\mathrm{m}}}-1\Big)^3\right),
\\
c_2&=-2V''_{\mathrm{m}}+\left(-\frac{3 R'''_{\mathrm{m}}V''_{\mathrm{m}}}{2R''_{\mathrm{m}}}-\frac{V'''_{\mathrm{m}}}{2}\right)(r_0-r_{\mathrm{m}})+O\left(\Big(\frac{r_0}{r_{\mathrm{m}}}-1\Big)^2\right),
\end{align}
where we have used $\lim_{r_0\to r_{\mathrm{m}}}(V'_0/R'_0)=V_{\mathrm{m}}''/R''_{\mathrm{m}}$. By inverting Eq.~\eqref{eq:bexp}, these expansions can be expressed in terms of $(b/b_{\mathrm{c}}-1)$ as follows:
\begin{align}
c_1&=4e^{-\lambda_{\mathrm{m}}}
R_{\mathrm{m}}R''_{\mathrm{m}}\left(
\frac{b}{b_{\mathrm{c}}}-1
\right)+O\left(
\Big(\frac{b}{b_{\mathrm{c}}}-1\Big)^{3/2}
\right),
\\
c_2&=-2V''_{\mathrm{m}}
-\frac{e^{-\lambda_{\mathrm{m}}/2}R_{\mathrm{m}}}{\sqrt{-V''_{\mathrm{m}}}}
\left(
\frac{3R'''_{\mathrm{m}}V''_{\mathrm{m}}}{R''_{\mathrm{m}}}+V'''_{\mathrm{m}}
\right)\Big(\frac{b}{b_{\mathrm{c}}}-1\Big)^{1/2}
+O\left(
\Big(\frac{b}{b_{\mathrm{c}}}-1\Big)
\right).
\end{align}
These results lead to the deflection angle in Eq.~\eqref{eq:alphab} with the following expression for the strong field limit coefficients:
\begin{align}
\bar{a}&=\sqrt{-\frac{2}{V''_{\mathrm{m}}}}, \\ 
\bar{b}&=-\bar{a}\log \frac{\bar{a}^2 R_{\mathrm{m}}R''_{\mathrm{m}}}{4 e^{\lambda_{\mathrm{m}}}}+ I_{\mathrm{R}}(r_\mathrm{m}) - \pi.
\end{align}
Thus, when $R'_{\mathrm{m}}=0$ (i.e., when the photon sphere coincides with the wormhole throat), the strong field limit coefficients arising from $I_{\mathrm{D}}$ can be expressed solely in terms of local quantities. In particular, $\bar{a}$ remains entirely determined by $V''_{\mathrm{m}}$ and retains the same form as in the $R'_{\mathrm{m}}\neq 0$ case, while $\bar{b}$ is given by
\begin{align}
\bar{b}&=-\bar{a}\log \left[\:\!
\frac{\bar{a}^2}{8}\left(1-2\:\! \frac{\mathrm{d}m}{\mathrm{d}R}\Big|_{r=r_{\mathrm{m}}}\right)
\:\!\right]+ I_{\mathrm{R}}(r_\mathrm{m}) - \pi. 
\end{align}
Here $\mathrm{d}m/\mathrm{d}R$ denotes the variation of the quasi-local mass $m$ with respect to the areal radius $R$.  Since $m$ is fundamentally defined as a function of the areal radius 
through the Misner--Sharp construction, $\mathrm{d}m/\mathrm{d}R$ provides an invariant measure of the local mass variation near the photon sphere.

\section{Formula for the deflection angle in the strong deflection limit}
\label{sec:4}
To facilitate a definite physical interpretation of the subsequent analysis, we introduce the following orthonormal basis: 
\begin{align}
e_{(0)}^a&=e^{-\nu/2} \:\!(\partial/ \partial t)^a,
\\
e_{(1)}^a&=e^{-\lambda/2}\:\!(\partial/\partial r)^a,
\\
e_{(2)}^a&=\frac{1}{R}(\partial/\partial \theta)^a,
\\
e_{(3)}^a&=\frac{1}{R \sin \theta}(\partial/\partial \varphi)^a.
\end{align}
Then, the tetrad components of the Einstein tensor $G_{(\mu)(\nu)}=G_{ab}e_{(\mu)}^a e_{(\nu)}^b$ are given by
\begin{align}
G_{(0)(0)}&=\frac{R'\lambda'}{R}e^{-\lambda}-\frac{(R')^2e^{-\lambda}-1}{R^2}-\frac{2 R''}{R}e^{-\lambda},
\label{eq:G00}
\\
G_{(1)(1)} &= \frac{R'\nu'}{R}e^{-\lambda}+\frac{(R')^2e^{-\lambda}-1}{R^2},
\label{eq:G11}
\\
G_{(2)(2)} &= G_{(3)(3)} = \frac{e^{-\lambda}}{2}\left[\:\!
\nu'' + \frac{(\nu')^2}{2} - \frac{\lambda'\nu'}{2} + R'\frac{\nu' - \lambda'}{R}+\frac{2R''}{R}
\:\!\right],
\label{eq:G22}
\end{align}
with all other components vanishing identically, where the prime denotes differentiation with respect to the radial coordinate $r$. These quantities $G_{(\mu)(\nu)}$ encode curvature as measured by static observers, thereby providing a coordinate-invariant characterization of the local geometry. The contracted Bianchi identity, $\nabla_{a}G^{ab}=0$, in the tetrad frame reads
\begin{align}
G_{(1)(1)}'+\frac{\nu'}{2}(G_{(0)(0)}+G_{(1)(1)})+\frac{2R'}{R}(G_{(1)(1)}-G_{(2)(2)})=0. 
\end{align}
Furthermore, from Eqs.~\eqref{eq:G00} and \eqref {eq:G11}, the derivatives of the metric functions can be expressed as
\begin{align}
\lambda' &= \frac{R e^\lambda}{R'} G_{(0)(0)}+\frac{(R')^2-e^\lambda}{RR'}+\frac{2R''}{R'}. 
\label{eq:lambda_prime}
\\
\nu' &= \frac{R e^\lambda}{R'} G_{(1)(1)}-\frac{(R')^2- e^\lambda}{RR'}.
\label{eq:nu_prime}
\end{align}
Solving Eq.~\eqref{eq:G22} for $\nu''$, we obtain
\begin{align}
\nu'' = 2e^\lambda G_{(2)(2)}
-\frac{(\nu')^2}{2} + \frac{\lambda' \nu'}{2} - R'\frac{\nu'- \lambda'}{R}-\frac{2R''}{R}.
\label{eq:nu2}
\end{align}

Now, we focus on $G_{(1)(1)}$ at $r=r_{\mathrm{m}}$ (assuming that the photon sphere exists). According to Eqs.~\eqref{eq:m} and \eqref{eq:PSradius}, this value is given by 
\begin{align}
G_{(1)(1)}^{\mathrm{m}}=\frac{2\left[\:\!R_{\mathrm{m}}-3m(r_{\mathrm{m}})\:\!\right]}{R_{\mathrm{m}}^3},
\label{eq:G11PS}
\end{align}
which establishes a direct link between the local geometrical quantity $G_{(1)(1)}^{\mathrm{m}}$ and the gravitational mass $m(r_{\mathrm{m}})$. Here, the value $G_{(1)(1)}^{\mathrm{m}}$ is positive when $R_{\mathrm{m}}>3m(r_{\mathrm{m}})$, negative when $R_{\mathrm{m}}<3m(r_{\mathrm{m}})$, and vanishes when $R_{\mathrm{m}}=3m(r_{\mathrm{m}})$. Moreover, $G_{(1)(1)}^{\mathrm{m}}$ is independent of the dynamical stability of photon circular orbits and remains valid even for anti-photon spheres~\cite{Gibbons:2016isj,Cunha:2017qtt,Kudo:2022ewn}. 

Similarly, we can relate $G_{(0)(0)}$ at $r=r_{\mathrm{m}}$ to the radial variation of the Misner--Sharp mass. Using Eq.~\eqref{eq:G00}, we have 
\begin{align}
G_{(0)(0)}^{\mathrm{m}}=\frac{2}{R^2_{\mathrm{m}}}\frac{\mathrm{d}m}{\mathrm{d}R}\bigg|_{r=r_{\mathrm{m}}}. 
\label{eq:G00PS}
\end{align}

Now, we establish the relation between $V''_{\mathrm{m}}$ and $G_{(\mu)(\mu)}^{\mathrm{m}}$, both evaluated at the photon sphere. Substituting Eqs.~\eqref{eq:PSradius}, \eqref{eq:lambda_prime}, and \eqref{eq:nu2} into Eq.~\eqref{eq:V''m} yields 
\begin{align}
V''_{\mathrm{m}}=-2\left[\:\!
1-R_{\mathrm{m}}^2 \left(G_{(0)(0)}^{\mathrm{m}}+G_{(2)(2)}^{\mathrm{m}}\right)
\:\!\right]. 
\label{eq:V''mG}
\end{align}
This expression demonstrates that the local spacetime geometry directly contributes to $V''_{\mathrm{m}}$. Notably, this contribution arises exclusively through the dimensionless product $R_{\mathrm{m}}^2 (G_{(0)(0)}^{\mathrm{m}}+G_{(2)(2)}^{\mathrm{m}})$, thereby underscoring its fundamental importance. In particular, a necessary condition for the existence of unstable photon circular orbits (i.e., $V_{\mathrm{m}}''<0$) is 
\begin{align}
R_{\mathrm{m}}^2\left(
G_{(0)(0)}^{\mathrm{m}}+G_{(2)(2)}^{\mathrm{m}}\right)< 1. 
\label{eq:PSC1}
\end{align}

Thus, using the results of Eqs.~\eqref{eq:G11PS}--\eqref{eq:V''mG}, the deflection angle in the strong deflection limit can be written as 
\begin{align}
\alpha(b) =-\bar{a} \log \left(
\frac{b}{b_{\mathrm{c}}}-1
\right)+\bar{b}+O\left(\left(
\frac{b}{b_{\mathrm{c}}}-1
\right)^{1/2}
\log\left(
\frac{b}{b_{\mathrm{c}}}-1
\right)
\right),
\end{align}
with the strong field limit coefficient $\bar{a}$ given by 
\begin{align}
\bar{a}&=\frac{1}{\sqrt{1-R_{\mathrm{m}}^2 \big(G_{(0)(0)}^{\mathrm{m}}+G_{(2)(2)}^{\mathrm{m}}\big)}},
\label{eq:newabar}
\end{align}
and the constant offset correction $\bar{b}$ given by
\begin{subequations}\label{eq:barb}
\begin{numcases}{\bar{b} =}
-\bar{a}\log\left[\frac{\bar{a}^2}{6}\left(1+R_{\mathrm{m}}^2G_{(1)(1)}^{\mathrm{m}}\right)\right]
+ I_{\mathrm{R}}(r_{\mathrm{m}}) - \pi, & if \(R'_{\mathrm{m}} \neq 0\), \label{eq:barb1} \\[2mm]
-\bar{a}\log\left[\frac{\bar{a}^2}{8}\left(1-R_{\mathrm{m}}^2 G_{(0)(0)}^{\mathrm{m}}\right)\right]
+ I_{\mathrm{R}}(r_{\mathrm{m}}) - \pi, & if \(R'_{\mathrm{m}} = 0\). \label{eq:barb2}
\end{numcases}
\end{subequations}
Notably, these expressions reveal that both the logarithmic divergence rate $\bar{a}$ and the constant offset correction $\bar{b}$ arising from $I_{\mathrm{D}}$ are determined solely by local geometrical quantities, i.e., $R_{\mathrm{m}}$ and $G_{(\mu)(\mu)}^{\mathrm{m}}$. Specifically, $\bar{a}$ depends only on $R_{\mathrm{m}}$ and the curvature sum $G_{(0)(0)}^{\mathrm{m}}+G_{(2)(2)}^{\mathrm{m}}$, which enters as the dimensionless product $R_{\mathrm{m}}^2(G_{(0)(0)}^{\mathrm{m}}+G_{(2)(2)}^{\mathrm{m}})$. Consequently, this combination essentially governs the rate of the logarithmic divergence of the strong deflection angle. 

The decomposition of $\bar{b}$ shows that its first term is determined solely by the dimensionless products $R_{\mathrm{m}}^2(G_{(0)(0)}^{\mathrm{m}}+G_{(2)(2)}^{\mathrm{m}})$ (through $\bar{a}$) and, depending on the case, either by $R_{\mathrm{m}}^2G_{(1)(1)}^{\mathrm{m}}$ for $R'_{\mathrm{m}}\neq0$ or by $R_{\mathrm{m}}^2G_{(0)(0)}^{\mathrm{m}}$ for $R'_{\mathrm{m}}=0$. Although, as discussed in the previous section, the decomposition between the divergent and regular parts depends on the choice of the expansion parameter $z$, the particular decomposition---with its first term expressed purely in terms of local quantities---remains highly valuable for estimating these local quantities. The remaining contribution to $\bar{b}$ arises from the regular part $I_{\mathrm{R}}(r_{\mathrm{m}})$, which is evaluated within a specific spacetime model. 

Consequently, these results imply that information about the curvature near the photon sphere can be extracted from measurements of the deflection angle in the strong deflection regime. Specifically, if $\bar{a}$ and $R_{\mathrm{m}}$ are determined from observations, the curvature combination $G_{(0)(0)}^{\mathrm{m}}+G_{(2)(2)}^{\mathrm{m}}$ can be directly derived. Moreover, if $\bar{b}$ is also measured and a specific spacetime model is assumed---so that all the relevant quantities can be computed for direct comparison with observations---then we can deduce $G_{(1)(1)}^{\mathrm{m}}$ [with $m(r_{\mathrm{m}})$] determined via Eq.~\eqref{eq:G11PS} in the case $R'_{\mathrm{m}}\neq0$, or $G_{(0)(0)}^{\mathrm{m}}$ (with $\mathrm{d}m/\mathrm{d}R|_{r=r_{\mathrm{m}}}$) determined via Eq.~\eqref{eq:G00PS} in the case $R'_{\mathrm{m}}=0$. In other words, if the theoretical predictions for $\bar{a}$ and $\bar{b}$ agree reasonably well with the observed values, this approach enables us to infer the local curvature and the gravitational mass, thereby providing a robust consistency check of the underlying gravitational model.

\section{Interpreting strong deflection in terms of local matter fields in general relativity}
\label{sec:5}
In this section, we focus on general relativity, in which the Einstein equations directly relate the Einstein tensor components to the local matter variables---the energy density $\rho(r)$, the radial pressure $P(r)$, and the tangential pressure $\Pi(r)$---as
\begin{align}
G_{(0)(0)}=8\pi \rho(r), \quad 
G_{(1)(1)}=8\pi P(r), \quad 
G_{(2)(2)}=G_{(3)(3)}=8\pi \Pi(r),
\label{eq:E-eqs}
\end{align}
where $\rho(r)$, $P(r)$, and $\Pi(r)$ denote the quantities measured by static observers.
In this discussion, we do not assume that the matter sector is a perfect fluid, and 
hence no specific equation of state is imposed on the quantities $\rho(r)$, $P(r)$, 
and $\Pi(r)$. The term ``matter" is used here in the standard 
general-relativistic sense to denote any contribution to the 
stress--energy tensor, including field configurations such as electromagnetic or 
scalar fields.

For clarity, we denote the matter field quantities evaluated at the photon sphere as
\begin{align}
\rho_{\mathrm{m}} \equiv \rho(r_{\mathrm{m}}), \quad
P_{\mathrm{m}} \equiv P(r_{\mathrm{m}}), \quad
\Pi_{\mathrm{m}} \equiv \Pi(r_{\mathrm{m}}).
\end{align}
Then, Eqs.~\eqref{eq:G11PS}--\eqref{eq:V''mG} can be expressed in terms of these matter field quantities as follows:
\begin{align}
P_{\mathrm{m}}&=\frac{R_{\mathrm{m}}-3m(r_{\mathrm{m}})}{4\pi R_{\mathrm{m}}^3},
\\
\rho_{\mathrm{m}}&=\frac{1}{4\pi R^2_{\mathrm{m}}}\frac{\mathrm{d}m}{\mathrm{d}R}\bigg|_{r=r_{\mathrm{m}}},
\\
V''_{\mathrm{m}}&=-2\left[\:\!
1-8\pi R_{\mathrm{m}}^2 \left(\rho_{\mathrm{m}}+\Pi_{\mathrm{m}}\right)
\:\!\right]. 
\label{eq:V''mG2}
\end{align}

We find that the inequality \eqref{eq:PSC1} for the existence of the photon sphere (i.e., $V_{\mathrm{m}}''<0$) can be rewritten as 
\begin{align}
8\pi R_{\mathrm{m}}^2 \left(\rho_{\mathrm{m}}+\Pi_{\mathrm{m}}\right)<1,
\end{align}
which implies that the sum $\rho_{\mathrm{m}}+\Pi_{\mathrm{m}}$ must be bounded from above for a photon sphere to exist. Notably, when $\rho_{\mathrm{m}}+\Pi_{\mathrm{m}}$ vanishes---as it does, for example, in vacuum in general relativity---the inequality is trivially satisfied, yielding $V''_{\mathrm{m}}=-2$. In this case, the result shows that a photon sphere can exist, whereas an anti-photon sphere cannot. 

Finally, we express the strong field limit coefficients $\bar{a}$ and $\bar{b}$ in terms of these matter field quantities. By substituting Eq.~\eqref{eq:E-eqs} into Eqs.~\eqref{eq:newabar} and \eqref{eq:barb}, we obtain
\begin{align}
\bar{a}&=\frac{1}{\sqrt{1-8\pi R_{\mathrm{m}}^2 \big(\rho_{\mathrm{m}}+\Pi_{\mathrm{m}}\big)}},
\label{eq:bara}
\end{align}
and
\begin{subequations}\label{eq:barb}
\begin{numcases}{\bar{b} =}
-\bar{a}\log\left[\frac{\bar{a}^2}{6}\left(1+8\pi R_{\mathrm{m}}^2 P_{\mathrm{m}}\right)\right] + I_{\mathrm{R}}(r_{\mathrm{m}}) - \pi, & for \(R'_{\mathrm{m}}\neq 0\), \label{eq:barb1} \\[2mm]
-\bar{a}\log\left[\frac{\bar{a}^2}{8}\left(1-8\pi R_{\mathrm{m}}^2 \rho_{\mathrm{m}}\right)\right] + I_{\mathrm{R}}(r_{\mathrm{m}}) - \pi, & for \(R'_{\mathrm{m}}=0\). \label{eq:barb2}
\end{numcases}
\end{subequations}
These expressions demonstrate that the divergent part contributions to the strong field limit coefficients are entirely determined by the local geometric scale $R_{\mathrm{m}}$ and by specific combinations of matter field variables. Specifically, $\bar{a}$ depends solely on $R_{\mathrm{m}}^2 \big(\rho_{\mathrm{m}}+\Pi_{\mathrm{m}}\big)$, while the first term in $\bar{b}$ is further influenced, via $\bar{a}$, by either $R_{\mathrm{m}}^2 P_{\mathrm{m}}$ (for $R'_{\mathrm{m}}\neq 0$) or $R_{\mathrm{m}}^2 \rho_{\mathrm{m}}$ (for $R'_{\mathrm{m}}= 0$). Based on the discussion at the end of Sec.~\ref{sec:3}, measurements of the deflection angle in the strong deflection regime allow us to infer the combination $\rho_{\mathrm{m}}+\Pi_{\mathrm{m}}$ and either the radial pressure $P_{\mathrm{m}}$ (for $R'_{\mathrm{m}}\neq 0$) or the energy density $\rho_{\mathrm{m}}$ (for $R'_{\mathrm{m}}= 0$), thereby directly linking observable lensing phenomena to the local matter distribution at the photon sphere.

In the following, we examine the relationship between the properties of the matter field quantities defined above and the strong field limit coefficients. Detailed calculations of the strong field limit coefficients for specific spacetime models have been extensively addressed in the literature, and we refer the reader to those works for further details. Nevertheless, it is instructive to present results for several well-known spacetimes. Furthermore, we elucidate the universal properties that emerge when specific combinations of matter-field quantities vanish.

As the simplest case, we consider a vacuum spacetime in the vicinity of the photon sphere. According to Birkhoff's theorem, this region is described by the Schwarzschild line element,
\begin{align}
\mathrm{d}s^2=-\left(1-\frac{2M}{r}\right)\mathrm{d}t^2+\left(1-\frac{2M}{r}\right)^{-1}\mathrm{d}r^2+r^2\:\!\mathrm{d}\Omega^2,
\end{align}
where $\mathrm{d}\Omega^2=\mathrm{d}\theta^2+\sin^2\theta\:\!\mathrm{d}\varphi^2$. Here, $M$ denotes the constant mass enclosed within the vacuum region, and the standard gauge $R=r$ is adopted. Since all matter field quantities vanish, at $r_{\mathrm{m}}=3M$ we obtain
\begin{align}
\rho(r_{\mathrm{m}})=0,\quad P(r_{\mathrm{m}})=0,\quad \Pi(r_{\mathrm{m}})=0.
\end{align}
Substituting these values into our expressions for the strong field limit coefficients yields
\begin{align}
\bar{a}=1, 
\quad
\bar{b}=\log 6+ I_{\mathrm{R}}(r_\mathrm{m}) - \pi,
\end{align}
with the critical impact parameter given by $b_{\mathrm{c}}=3\sqrt{3}M$. 
Note that $\bar{a}=1$ is a direct consequence of the vanishing matter fields in a vacuum spacetime. These results recover the well-known expressions for the Schwarzschild solution~\cite{Bozza:2002zj, Tsukamoto:2016jzh} because our definition of $z$ is equivalent to that used in those references for the vacuum case. This confirms that our formulation is fully consistent with the standard vacuum case in general relativity.

Next, we consider an electrovacuum spacetime (i.e., a vacuum spacetime containing only electromagnetic fields) in the vicinity of the photon sphere. According to Birkhoff's theorem, this region is described by the Reissner--Nordstr\"om 
line element,
\begin{align}
\mathrm{d}s^2=-\left(1-\frac{2M}{r}+\frac{Q^2}{r^2}\right)\mathrm{d}t^2+\left(1-\frac{2M}{r}+\frac{Q^2}{r^2}\right)^{-1}\mathrm{d}r^2+r^2\:\!\mathrm{d}\Omega^2,
\end{align}
where $M$ and $Q$ denote the constant mass and electric charge characterizing the electrovacuum region, respectively, and the standard gauge $R=r$ is adopted. In the electrovacuum, matter field quantities are given by
\begin{align}
\rho(r)=\frac{Q^2}{8\pi r^4}, \quad P(r)=-\frac{Q^2}{8\pi r^4}, \quad \Pi=\frac{Q^2}{8\pi r^4}, 
\end{align}
Using these expressions, we obtain the strong field limit coefficients
\begin{align}
\bar{a}=\frac{1}{\sqrt{1-2Q^2/r_{\mathrm{m}}^2}},\quad
\bar{b}=-\bar{a} \log\left[\:\!
\frac{\bar{a}^2}{6}\left(1-Q^2/r_{\mathrm{m}}^2\right)
\:\!\right]+ I_{\mathrm{R}}(r_\mathrm{m}) - \pi,
\end{align}
where $r_{\mathrm{m}}=(3M+\sqrt{9M^2-8Q^2})/2$ and $b_{\mathrm{c}}^2=3r_{\mathrm{m}}^2/(1-Q^2/r_{\mathrm{m}}^2)$. 
Note that these results agree with those in Refs.~\cite{Eiroa:2002mk,Tsukamoto:2016oca}, which employ the same expansion parameter as used here. These expressions indicate that both coefficients depend solely on the dimensionless parameter $Q/r_{\mathrm{m}}$. 

Next, consider a special case in which the spacetime is nonvacuum, but matter fields satisfy $\rho(r_{\mathrm{m}})+\Pi(r_{\mathrm{m}})=0$. In this situation, the strong field limit coefficient $\bar{a}$ remains equal to $1$, thereby mirroring the Schwarzschild result despite the presence of nonvanishing matter fields. This demonstrates that, when $\rho(r_{\mathrm{m}})+\Pi(r_{\mathrm{m}})=0$, the strong deflection limit exhibits universal behavior with $\bar{a}=1$, irrespective of the specific details of the matter content.

As a specific example, we consider a massless scalar field $\Phi$ minimally coupled to gravity. Its stress-energy tensor is given by (see, e.g., Ref.~\cite{Herdeiro:2015waa})
\begin{align}
T_{ab}=(\nabla_a \Phi)( \nabla_b \Phi)-\frac{1}{2} g_{ab}\:\! g^{cd} (\nabla_c \Phi)( \nabla_d \Phi).
\end{align}
For a static, spherically symmetric scalar field $\Phi(r)$ in a static, spherically symmetric spacetime, we find
\begin{align}
\rho(r) = P(r) = -\Pi(r)=\frac{e^{-\lambda}}{2} (\Phi')^2. 
\end{align}
This result explicitly shows that $\rho+\Pi=0$ throughout the spacetime, which naturally leads to a universal behavior in the strong deflection limit by ensuring that $\bar{a}=1$. In other words, even though the spacetime is not strictly vacuum due to the presence of the massless scalar field, the specific structure of its stress-energy tensor guarantees that $\bar{a}=1$, thereby retaining the Schwarzschild value.

This analysis clearly explains the property $\bar{a}=1$, which, although consistently observed in many disparate examples, had remained unexplained in terms of its fundamental physical origin. One notable example is the Janis--Newman--Winicour (JNW) spacetime~\cite{Janis:1968zz,Wyman:1981bd}, which is described by the line element
\begin{align}
\mathrm{d}s^2=-\left(1-\frac{r_{\mathrm{g}}}{r}\right)^\gamma \mathrm{d}t^2+\left(1-\frac{r_{\mathrm{g}}}{r}\right)^{-\gamma}\mathrm{d}r^2
+\left(1-\frac{r_{\mathrm{g}}}{r}\right)^{1-\gamma} r^2\:\!\mathrm{d}\Omega^2,
\end{align}
where $\gamma$ and $r_{\mathrm{g}}$ are constant parameters. For $1/2<\gamma<1$, this solution describes weakly naked singularities at $r=r_{\mathrm{g}}$, accompanied by a nontrivial scalar field profile, and features the photon sphere located at $r=r_{\mathrm{g}}(2\gamma+1)/2$. The associated scalar field quantities are given by 
\begin{align}
\rho(r)=P(r)=-\Pi(r)=\frac{r_\mathrm{g}^2\left(
1-r_{\mathrm{g}}/r
\right)^\gamma(1-\gamma^2)
}{32\pi r^2(r-r_{\mathrm{g}})}.
\end{align}
Since this matter configuration satisfies $\rho+\Pi=0$ throughout the spacetime, the strong deflection limit coefficients are given by
\begin{align}
\bar{a}=1, \quad \bar{b}=\log \left[\:\!
\frac{2(2\gamma+1)(2\gamma-1)}{\gamma^2}
\:\!\right]+ I_{\mathrm{R}}(r_\mathrm{m}) - \pi,
\end{align}
where $b_{\mathrm{c}}=(r_{\mathrm{g}}/2)(2\gamma+1)^{(2\gamma+1)/2}(2\gamma-1)^{-(2\gamma-1)/2}$. Thus, even in the presence of a nontrivial scalar field, if $\rho+\Pi=0$, the strong deflection limit exhibits the universal behavior characterized by $\bar{a}=1$. For completeness, note that the first term in $\bar{b}$ differs from that in Refs.~\cite{Bozza:2002zj,Chen:2023uuy} due to a different choice of the parameter $z$.

Another notable example is the Ellis--Bronnikov wormhole spacetime~\cite{Ellis:1973yv, Bronnikov:1973fh}, which is given by the line element
\begin{align}
\mathrm{d}s^2=-\mathrm{d}t^2+\mathrm{d}r^2+(r^2+a^2) \mathrm{d}\Omega^2,
\end{align}
where $a$ is a constant parameter. This solution describes a traversable wormhole with its throat located at $r=0$ (i.e., the areal radius $R=a$), supported by a phantom massless scalar field, and it possesses the photon sphere at $r=0$. Therefore, $R_{\mathrm{m}}'=0$ holds in this case. 
The associated scalar field quantities are given by 
\begin{align}
\rho(r)=P(r)=-\Pi(r)=-\frac{a^2}{8\pi (r^2+a^2)^2}. 
\end{align}
This matter configuration satisfies $\rho+\Pi=0$, thereby leading to $\bar{a}=1$; and furthermore, since $R'_{\mathrm{m}}=0$, the coefficient $\bar{b}$ is determined by Eq.~\eqref{eq:barb2}. Finally, we obtain
\begin{align}
\alpha(b) &=-\bar{a} \log \left(
\frac{b}{b_{\mathrm{c}}}-1
\right)+\bar{b}+O\left(\left(
\frac{b}{b_{\mathrm{c}}}-1
\right)
\log\left(
\frac{b}{b_{\mathrm{c}}}-1
\right)
\right),
\label{eq:alphaEB}
\\
\bar{a}&=1, \quad \bar{b}=2\log 2+ I_{\mathrm{R}}(r_\mathrm{m}) - \pi,
\end{align}
which agree with those in Ref.~\cite{Tsukamoto:2016qro}, employing the same expansion parameter used here. Note that for the Ellis--Bronnikov wormhole under consideration, the conditions $R'''_{\mathrm{m}}=0$ and $V'''_{\mathrm{m}}=0$ are satisfied, which causes the coefficient of $(b/b_{\mathrm{c}}-1)^{1/2}$ in $c_2$ to vanish. Consequently, the error term in Eq.~\eqref{eq:alphaEB} is of order $(b/b_{\mathrm{c}}-1)\log(
b/b_{\mathrm{c}}-1)$.

\section{Summary and discussion}
\label{sec:6}
In this paper, we have proposed a novel formulation of the deflection angle in the strong deflection limit for static, spherically symmetric spacetimes. By introducing an expansion parameter based on the areal radius---which has clear geometric significance---and isolating the logarithmically divergent part of the integral, we have derived the strong field limit coefficients that characterize the divergence rate and the constant offset correction. While previous studies expressed these coefficients in a metric- and coordinate-dependent manner, our formulation relates them to the second derivative of the effective potential, which governs photon dynamics near the photon sphere, and to the photon sphere radius and the quasi-local mass. Furthermore, by relating these quantities to the tetrad components of the Einstein tensor, we have demonstrated that the strong field limit coefficients depend on local, coordinate-invariant curvatures. Since these curvatures are connected to matter distribution quantities via the Einstein equations, our results clarify how the logarithmic divergence of the deflection angle in the strong deflection limit is determined by matter field quantities. Note that since our results are expressed in terms of the Einstein tensor, they are, in essence, independent of the underlying gravitational theory.

Our new formula demonstrates that the logarithmic divergence rate, $\bar{a}$, is determined by the product of the photon sphere's areal radius squared, $R_{\mathrm{m}}^2$, and the sum of the specific tetrad components of the Einstein tensor, $G_{(0)(0)}^{\mathrm{m}}+G_{(2)(2)}^{\mathrm{m}})$, evaluated at the photon sphere. This result indicates that $R_{\mathrm{m}}^2 (G_{(0)(0)}^{\mathrm{m}}+G_{(2)(2)}^{\mathrm{m}})$ is the fundamental factor characterizing the strong deflection phenomenon. Within the framework of general relativity, the Einstein equations ensure that these local curvature quantities correspond to the local energy density and tangential pressure (i.e., $\rho_{\mathrm{m}}$ and $\Pi_{\mathrm{m}}$), implying that $\bar{a}$ depends solely on $R_{\mathrm{m}}^2\left(\rho_{\mathrm{m}}+\Pi_{\mathrm{m}}\right)$. Therefore, by precisely measuring the logarithmic divergence rate of the deflection angle, we can directly determine $G_{(0)(0)}^{\mathrm{m}}+G_{(2)(2)}^{\mathrm{m}}$, or, assuming general relativity, $\rho_{\mathrm{m}}+\Pi_{\mathrm{m}}$. Moreover, the constant offset correction, $\bar{b}$, also encodes information about the local curvature and the quasi-local mass; thus, a comprehensive evaluation of both coefficients is expected to yield deep insights into the internal structure and local physical conditions of compact objects. 

Our formulation clearly explains why the special value $\bar{a}=1$ is ubiquitously observed across diverse spacetime models. In particular, at the photon sphere, if the matter fields satisfy $\rho_{\mathrm{m}}+\Pi_{\mathrm{m}}=0$, then $R_{\mathrm{m}}^2\left(\rho_{\mathrm{m}}+\Pi_{\mathrm{m}}\right)$ vanishes, immediately implying $\bar{a}=1$. This insight crucially resolves the long-standing puzzle regarding the behavior of the logarithmic divergence rate.

Furthermore, using our results, we show that, within the eikonal approximation, the QNM frequencies can be expressed in terms of the local geometric and material quantities at the photon sphere. The real part of the QNM frequencies is related to the orbital frequency at the photon sphere, while the imaginary part is determined by the Lyapunov exponent of the unstable photon circular orbit. In particular, we obtain
\begin{align}
\omega_{\mathrm{QNM}}=\Omega_{\mathrm{c}} \:\! l-i\left(n+\frac{1}{2}\right)|\lambda_{\mathrm{L}}|,
\end{align}
where $n$ is the overtone number, $l$ is the angular momentum of the perturbation, and $\Omega_{\mathrm{c}}=1/b_{\mathrm{c}}$ denotes the angular coordinate velocity of the unstable photon circular orbit. 
The Lyapunov exponent is given by $\lambda_{\mathrm{L}}=b_{\mathrm{c}}^{-1}\sqrt{-V''_{\mathrm{m}}/2}$, which is related to $\bar{a}$ as~\cite{Stefanov:2010xz,Raffaelli:2014ola}
\begin{align}
\lambda_{\mathrm{L}}=\frac{1}{b_{\mathrm{c}}\bar{a}}.
\end{align}
Therefore, by applying our formula, we can express $\lambda_{\mathrm{L}}$ as 
\begin{align}
\lambda_{\mathrm{L}}=\frac{\sqrt{1-8\pi R_{\mathrm{m}}^2 (\rho_{\mathrm{m}}+\Pi_{\mathrm{m}})}}{b_{\mathrm{c}}}.
\end{align}
Thus, the QNM frequencies are ultimately determined by the same local matter field (or local curvature) quantities that govern the deflection angle.%
\footnote{It should be noted that the correspondence between photon spheres and eikonal QNMs is not strictly universal; for example, in Lovelock gravity, the eikonal QNM frequencies of gravitational perturbations deviate from the photon-sphere prediction due to the spin-dependent structure of the effective potential~\cite{Konoplya:2017wot}.}
Therefore, precise measurements of QNM frequencies---such as those from gravitational wave observations---provide an independent means to probe the local curvature and matter distribution near the photon sphere, thereby reinforcing the connection between strong gravitational lensing and the dynamical properties of compact objects. Furthermore, our findings may also shed light on a universal upper bound on chaos in thermal quantum field theory~\cite{Maldacena:2015waa,Hashimoto:2016dfz}, which provides an inequality between the Lyapunov exponent and the surface gravity of the horizon, a relation whose generalization to the photon sphere has recently been discussed~\cite{Gallo:2024wju}.

Our formulation is a universal, coordinate-independent framework that applies to a broader range of spacetime models than conventional approaches, promising numerous future applications. Moreover, when expressed in terms of the Einstein tensor, the formulation becomes independent of any specific gravitational theory, allowing its application to solutions beyond general relativity. Consequently, further investigations are required to clarify how the strong field limit coefficients depend on matter fields in modified gravitational theories, and our formulation is expected to serve as a foundation for clarifying their physical significance. Although the present study focuses on static, spherically symmetric spacetimes, extending the formulation to less symmetric cases remains an important direction for future research. In addition, applying our formulation to the strong deflection of massive particles offers another promising avenue for further investigation (for perturbations of the stable circular orbit of massive particles, see Ref.~\cite{Harada:2022uae}). Furthermore, since a relationship between QNM frequencies and local curvature and matter fields is suggested, combining high-precision measurements of quasinormal modes from gravitational wave observations with analyses based on our formulation is expected to set the stage for a novel approach that yields detailed insights into the internal structure and local physical conditions of compact objects in extreme regimes.

\begin{acknowledgments}
The author gratefully acknowledges useful discussions and valuable comments from Ken-ich Nakao, Tomohiro Harada, Tsutomu Kobayashi, Yasuyuki Hatsuda, Hideki Ishihara, Chulmoon Yoo, Akio Hosoya, Yohsuke Takamori, Koji Nakamura, Masashi Kimura, Tetsuya Shiromizu, Hiroshi Kozaki, Shinya Aoki, and Naoki Tsukamoto. This work was supported in part by JSPS KAKENHI Grants No.~JP22K03611, No.~JP23KK0048, and No.~JP24H00183 and by Gakushuin University.
\end{acknowledgments}

\end{document}